\documentclass[11pt,letterpaper]{article}
\usepackage{enumitem}
\usepackage[margin=1in]{geometry} % full-width

\usepackage{amsmath}
\usepackage{amsthm}
\usepackage{amssymb}
\usepackage{bm}
\usepackage{cancel}
\usepackage[normalem]{ulem}
\usepackage{setspace}
\usepackage{comment}
\usepackage[utf8]{inputenc}
\usepackage{hyperref}
\hypersetup{
	unicode,
	pdfauthor={Raymond J. Hinton, Jr. and Pepa Ramírez Cobo and Brani Vidakovic},
	pdftitle={The Dual Wavelet Spectra: An Alternative Perspective on Hurst Exponent Estimation with Application to Mammogram Classification},
	pdfsubject={wavelets, statistics},
	pdfkeywords={scaling, non-decimated wavelet transform, Hurst exponent, 2D fractional Brownian motion, image classification},
	pdfproducer={LaTeX},
	pdfcreator={pdflatex}
}

\usepackage[sort&compress,numbers,square]{natbib}
\theoremstyle{plain}

\theoremstyle{definition}

\newcommand \bbE{\mathbb{E}}

\usepackage{subcaption}
\usepackage{booktabs}
\usepackage{tabularx}
\usepackage{multirow}
\usepackage{graphicx, color}
\graphicspath{{fig/}}

\usepackage{algorithm, algpseudocode} % use algorithm and algorithmicx for typesetting algorithms
\usepackage{mathrsfs} % for \mathscr command

\usepackage{lipsum}

\title{The Dual Wavelet Spectra: An Alternative Perspective on Hurst Exponent Estimation with Application to Mammogram Classification}
\author{Raymond J. Hinton, Jr.$^a$ \and Pepa Ram{\'i}rez Cobo$^b$ \and Brani Vidakovic$^a$}
\date{
	$^a$Department of Statistics, Texas A\&M University, College Station, TX, USA \\ \texttt{\{rhinton, brani\}@tamu.edu}\\%
	$^b$Department of Statistics and Operations Research, University of Cádiz, Cádiz, Spain \\ \texttt{pepa.ramirez@uca.es}\\[2ex]%
}

\begin{document}

\maketitle

\begin{abstract} 

The wavelet spectra is a common starting point for estimating the Hurst exponent of a self-similar signal using wavelet-based techniques. The decay of the $\log_2$ average energy of the detail wavelet coefficients as a function of the level of signal decomposition can be used to construct estimators for this parameter. In this paper, we expand on previous work which introduced the ``dual" wavelet spectra, where decomposition levels are instead treated as a function of energy values, and propose a relationship between its slope and the Hurst exponent by inverting the standard wavelet spectra, thereby creating a new estimator. The effectiveness of this estimator and its sensitivity to several settings are demonstrated through a simulation study. Finally, we show how the technique performs as a feature extraction method by applying it to the task of detecting the presence of breast cancer in mammogram images. Dual spectra wavelet features had a statistically significant effect on the log-odds of Cancer.
\\
\noindent\textbf{Keywords:} scaling, non-decimated wavelet transform, Hurst exponent, 2D fractional Brownian motion, image classification
\end{abstract}

\section{Introduction}
\label{sec:intro}

Many processes in nature such as turbulence, tree growth, ocean waves, and brain signals exhibit complex patterns. Self-similarity is a property of many such phenomena that can summarize important characteristics of their underlying dynamics. In the context of signal processing, self-similarity refers to the repetitive structures at different resolutions of a signal. This property can be used to extract meaningful features for tasks such as pattern recognition, fractal analysis, and image compression. Thus, statistical methodologies that measure self-similarity can offer deeper insight about signals beyond traditional summaries.

The Hurst exponent is a parameter that quantifies the self-similarity in a signal. A number of estimators have been proposed for the Hurst exponent, including the rescaled range method of \citep{hurst1951, mandelbrot1968}, and others as reviewed in \citep{coeurjolly2000}. Techniques based on wavelets have been particularly effective. Wavelets decompose a signal at a number of levels, or scales, making them a natural tool for investigating self-similarity. A standard technique to estimate the Hurst exponent uses the slope of the wavelet spectra, which is a plot of the logarithm of the wavelet energies (squared absolute values of detail wavelet coefficients) as a function of the wavelet decomposition level \citep{flandrin1992wavelet,veitchabry1999,abry1998wavelet}. Extensions to this technique include correcting the bias from estimating the slope \citep{abry1998wavelet}, estimating wavelet energies with a robust distance covariance approach \citep{vimalajeewa2023early}, considering higher dimensions \citep{Nicolis2011,ramirez-cobo2011wavelet, Hamilton2024waveletscaling}, or using the richer wavelet packet decomposition to analyze the signal \citep{jones1996wavelet, wang2006estimating, hinton2025ovarian}, among other approaches.

Researchers have also gone beyond the wavelet spectra approach to estimate the Hurst exponent with other wavelet-based techniques. For example, \citep{vimalajeewa2025advanced} calculates exact distributional properties of a function of pairs of wavelet energies, and aggregates these pairwise estimators to yield a robust estimate of the Hurst exponent. In this work, we describe an alternative to the ``primal" wavelet spectra\footnote{We use ``primal" to distinguish the usual wavelet spectra from the ``dual" spectra proposed herein.}, introduced as the ``dual" wavelet spectra in \citep{kong2019assessing}, where instead the average levels of the wavelet decomposition are considered as a function of the wavelet energies. To our knowledge, \cite{Jensen1999} first and \cite{roux2004dual} later constitute some of the first works investigating dual formulations of scaling relationships. They both explore \emph{inverted} structure functions in the context of turbulence.

The main contribution of this work is to provide a novel wavelet-based spectral method for estimating the Hurst exponent in 2D fractional Brownian motion. We carry out simulation studies to understand the method's sensitivities to different parameters and its performance at different values of the Hurst exponent. To explore whether the alternative perspective quantifies additional information beyond the primal wavelet spectra in complex real-world data, we compare the classification performance of dual and primal wavelet spectra-based features for detecting breast cancer in mammogram images. Features calculated from our estimator have a statistically significant effect on the log-odds of cancer, providing evidence that the method captures valuable information for use in classification problems of images presenting different scaling patterns. 

The remainder of this paper is organized as follows. In Section \ref{sec:prelim}, we provide background on wavelet transforms, self-similarity, and the primal wavelet spectra. Section \ref{sec:dual} defines the dual wavelet spectra, derives an estimator for the Hurst exponent of a 2D fractional Brownian motion, and provides results of a simulation study using the estimator. Section \ref{sec:application} describes how the technique is used to extract features from mammograms to use for classification. Finally, Section \ref{sec:concl} summarizes the work and contains ideas for future research.

%%%%%%%%%%
\section{Preliminaries} \label{sec:prelim}

As we will show in Section \ref{sec:dual}, the dual wavelet spectra for 2D fractional Brownian motion is found using the non-decimated wavelet transform (NDWT) of a signal. Since the dual wavelet spectra is based inverting the primal wavelet spectra, in this section we provide an overview of wavelet transforms, self-similarity, fractional Brownian motion, and primal wavelet spectra for 2D signals (e.g. images).

\subsection{Wavelet transforms}

A wavelet transform decomposes a signal at multiple resolutions (also referred to as scales or levels). The transformed signal comprises one level of \textit{detail} wavelet coefficients for each level of decomposition, and a single level of \textit{approximation} wavelet coefficients (hereafter, just ``detail" and ``approximation" coefficients, respectively). For brevity and consistency, we consider signals that have dimensions which are a power of 2, i.e., vectors $N \times 1$ or matrices $N \times N$, where $N = 2^J$ for some $J \in \mathbb{Z}^+$. Signals with dimensions that are not a power of 2 can be either adapted (e.g. zero-padding) or transformed directly with certain wavelet transforms. The first level of decomposition, indexed $j = J-1$, captures fine details in the original signal. The next level, $J-2$, captures coarser details, and the level indices decrease\footnote{Some authors use a convention where level indices start at 1 and \textit{increase} as the levels become more coarse.} to a chosen coarsest detail level $J_0 = J - L$, where $L$ is the number of detail levels in the decomposition. 

The discrete wavelet transform (DWT) for 1D and 2D signals begins with choosing a  wavelet function, $\psi$, and a number of detail levels, $L$. These choices define a transform matrix, $W$ (which is constructed differently in the following cases, but notated the same for the sake of simplicity). The 1D DWT of a signal $Y$ is $X = WY$, where the vector $X$ has the same length as $Y$. The first level of detail coefficients is half the length of the original signal, and each subsequent level is half the length of the preceding level. The 2D DWT of a square matrix $Y$ is similarly $X = WYW^\dagger$, where $X$ is a matrix that is the same size as $Y$ ($W^\dagger$ is the Hermitian transpose of $W$, to accommodate complex-valued wavelets). In this context, $X$ contains multiple levels of horizontal, vertical, and diagonal detail coefficient regions, and a single region of approximation coefficients. Instead of using matrix multiplication, these transformations can also be performed using Mallat's algorithm \cite{mallat1989pyramid}: taking one level of decomposition of the approximation coefficients (using the original signal, first) and discarding half (usually, every-other) of the resulting detail and approximation coefficients before repeating these steps. Discarding half of the coefficients is called \textit{decimation}, and leads to the decomposed signal being the same size as the original signal. 

The NDWT (see, for example \cite{Shensa1992,NasonSilverman1995}) is a different class of wavelet transforms that does not perform the decimation step at each level. It presents some advantages over the standard discrete wavelet transform. For example, NDWT produces a redundant representation which improves robustness against noise and increases stability. In addition, NDWT avoids downsampling, so the coefficients vary smoothly with signal shifts. In this work, we use the \textit{scale-mixing} 2D NDWT, which was introduced as an extension of the classic 2D separable wavelet transform, where both axes are refined by the same scale at each decomposition level. In the scale-mixing case, different scales are allowed along different directions compared to the standard 2D NDWT. This makes it better at capturing anisotropic and directional features (like textures or elongated edges), and it achieves faster approximation rates for functions with mixed smoothness.  For more detail regarding scale-mixing wavelet transforms, see \cite{ramirez-cobo2011wavelet,Remenyi2014}. For the purposes of this work, all of the described results hold for both the scale-mixing and standard 2D NDWT, since the diagonal region detail coefficients are the same for both. The scale-mixing 2D NDWT is computed as $X = W_1 Y W_2^\dagger$ for a rectangular matrix $Y$ (see \cite{kang2016NDWTscalemixing} for more details about the transform matrices $W_1$ and $W_2$). The matrix $X$ contains a number of regions that are the same size as the original signal -- most crucially, one region of diagonal detail coefficients for each level of decomposition. The decomposition region can be specified by a pair $\bm{j} = (j_1, j_2)$, and within a region, a specific coefficient can be specified by a pair $\bm{k} = (k_1, k_2)$. We concentrate on detail coefficients, denoted with $d$, and so a specific detail coefficient is specified by $d_{\bm j; \bm k}$. Coefficients in diagonal decomposition regions, where $\bm{j} = (j, j)$ for some decomposition level $j$, will be used in the primal and dual spectra. 

Additional information about the properties and implementation of wavelet transforms can be found in monographs such as \cite{daubechies1992tenlectures, mallat1999wavelettour, vidakovic2009statisticalwavelets}.

\subsection{Self-similarity and primal wavelet spectra}\label{sec:self-sim}

A 2D stochastic process $\{ X(\bm{t}), \bm{t} \in [0,1] \times [0,1] \}$ is self-similar with Hurst exponent $H$ if, for any $a \in \mathbb{R}^+$,
$
X(a \bm{t}) \overset{d}{=} a^H X(\bm{t}),
$
where $\overset{d}{=}$ denotes equality of all joint finite-dimensional distributions \citep{Hamilton2024waveletscaling}. If, in addition, $X(\bm{t})$ is Gaussian with stationary zero-mean increments, then $X(\bm{t})$ is a two-dimensional fractional Brownian motion (2D fBm), or a fractional Brownian field (fBf). It is known \citep{riedi1995spectral} that the autocorrelation function of $X(\bm{t})$ is given by
\begin{equation}
    \bbE \left[X(\bm{t})X(\bm{s})\right]=\frac{\sigma^2_H}{2}\left(||\bm{t}||^{2H}+||\bm{s}||^{2H}-||\bm{t}-\bm{s}||^{2H}\right),
\end{equation}
where $||\cdot||$ is the usual Euclidean norm in $\mathbb{R}^2$ and 
\begin{equation}
    \sigma^2_H = \frac{2^{-(1+2H)}\Gamma(1-H)}{\pi H \Gamma(1+H)}.
\end{equation}
For the DWT and NDWT of 2D fBm, it can be shown that the expectation of the squared absolute value (also called the \textit{energy)} of a diagonal detail coefficient in level $j$ is
\begin{equation}
    \bbE \left( |d_{(j,j); \bm{k}}|^2 \right) = \frac{\sigma_H^2}{2} V_\psi 2^{-(2 H +2) j},
\end{equation}
where $V_\psi$ is a constant depending only on the wavelet $\psi$ and exponent $H$, but not on the level $j$. For brevity, we will use ``wavelet energy" to specifically refer to the energy of diagonal detail coefficients, since we only use these coefficients in this work. For the result, see \cite{Nicolis2011} for the 2D DWT, and \cite[eq. (4)]{kang2022NDWT2dspectrum} for the 2D NDWT. Taking the base 2 logarithm of both sides yields
\begin{equation}
\log_2 \bbE \left(|d_{(j,j); \bm{k}}|^2 \right) = -(2H + 2) j + C_{\psi, H}, \label{eq:primal-linear}
\end{equation}
for some constant $C_{\psi, H}$ independent of $j$. For the sake of distinction, we will call (\ref{eq:primal-linear}) the \textit{theoretical primal wavelet spectra}. The \textit{empirical primal wavelet spectra} (hereafter, \textit{primal spectra}, for brevity) comprises pairs of levels $j$ and estimates of the $\log_2$ expectation in (\ref{eq:primal-linear}). That is, pairs $
\left(j, \log_2 \overline {|d_{(j,j); \bm{k}}|^2} \right),$
where 
\begin{equation}
\overline {|d_{(j,j); \bm{k}}|^2} = ({1}/{n_{(j,j)}})\sum_{(k_1, k_2)} |d_{(j,j); (k_1, k_2)}|^2
\end{equation}
is the sample mean of wavelet energies in detail region $(j,j)$, and $n_{(j,j)}$ is the number of coefficients in the region.
In practice, a subset of the levels $\{ j_1, \ldots,  j_N\}$ may be chosen where the primal spectra is linear. Let $\beta_p = -(2H +2)$ be the slope of the primal spectra. Once an estimate $\hat \beta_p$ of the slope is found, the Hurst exponent can be estimated as $\hat H_p = -\hat \beta_p/2 - 1$.

\section{Dual wavelet spectra} \label{sec:dual}

In this section, we present a derivation for the empirical dual wavelet spectra which essentially involves inverting the linear relationship for the empirical primal wavelet spectra. Although our approach is informal, we find that it can perform comparably to primal spectra estimators in simulation studies. This idea is inspired by earlier works such as \cite{Jensen1999} and \cite{roux2004dual}, in which dual versions of fractal spectra are calculated. The following derivation can be applied to the 2D DWT, standard 2D NDWT, and scale-mixing 2D NDWT, although in practice, we use the scale-mixing 2D NDWT as will be explained in Section \ref{sec:implement}.

\subsection{Derivation}

The primal spectra can be summarized as a linear regression of $\log_2$ wavelet energies on level indices. We propose formulating the \textit{empirical dual wavelet spectra} (hereafter, \textit{dual spectra}, for brevity) as a linear regression of level indices on $\log_2$ wavelet energies. In particular, the points of the primal spectra are $\log_2$ sample mean wavelet energies with given levels. Analogously, we will compute points in the dual spectra using the average levels of detail coefficients with given $\log_2$ wavelet energies. Let $e \in [0, \infty)$ be some wavelet energy value. We introduce this notation to distinguish between an arbitrary wavelet energy value across all levels and actual observed wavelet energies in the level $j$, $|d_{(j,j);\bm k}|^2$.  For comparison, the points of the two spectra can be written as:
\begin{enumerate}[label=(\roman*)]
    \item Primal spectra: $(\log_2 \overline {|d_{(j,j);\bm k}|^2}, j)$ - $\log_2$ average wavelet energy and the specific level $j$,
    \item Dual spectra: $(\bar j, \log_2 e)$ - average level, $\bar j$, for detail coefficients with a specific $\log_2$ wavelet energy, $\log_2 e$.
\end{enumerate}

Although it is conceptually simple to arrive at this form of the dual spectra from the primal version, this will not work in practice since the $\log_2 e$ values are continuous, not discrete. For any specific value of $\log_2 e$, we are likely to have at most one detail coefficient with $\log_2 |d_{(j,j); \bm k}|^2 = \log_2 e$, and therefore at most one detail coefficient with which to calculate the corresponding $\bar j$. Thus, we group detail coefficients in intervals of $\log_2 e$ values, which will ensure multiple coefficients are available for calculating $\bar j$. This grouping will be done using empirical quantiles of the observed wavelet energy values. In the rest of this section, we will derive the equation for the dual spectra, describe the selection of intervals of $\log_2 e$ values, calculate the mean level $j$ for an interval, rewrite the dual spectra in terms of these intervals and mean levels, and finally introduce the estimator for $H$ based on the dual spectra.

From (\ref{eq:primal-linear}), the relationship between wavelet energies and levels in the primal spectra is 
\begin{equation}
{\log_2 \overline {|d_{(j,j);\bm k}|^2} \approx \allowbreak -(2H + 2)j + C_{\psi,H}.}    
\end{equation}
We propose finding the equation for the dual spectra by replacing the sample mean with a single arbitrary wavelet energy, $e$, replacing $j$ with $\bar j$ (the average level of detail coefficients with this $\log_2 e$), and solving for $\bar j$. This leads to
\begin{equation}
    \bar j = -[1 / (2H + 2)] \log_2 e + C'_{\psi, H},
\end{equation}
where $C'_{\psi, H}$ is a new constant that depends on the mother wavelet and $H$, but not $e$.

Now we construct intervals of $\log_2 e$ values from which we can calculate mean levels.
The intervals, $I_m$ for $i = 1, ..., M$, will be bounded by the quantiles of the observed $\log_2$ wavelet energies to ensure each interval has an even split of data points. Specifically, we choose the following $M-1$ empirical quantiles $q_i$ to determine $M$ intervals as follows:
\begin{equation}
    \left| \left\{(j, \bm k) : |d_{(j,j);\bm k}|^2 \leq q_i \right\} \right| = 
    \frac {i - 0.5} {M-1}, \quad i = 1, \ldots, M-1.
\end{equation}
Let $q_0 = \min (|d_{(j,j);\bm k}|^2)$ and  $q_M = \max ( |d_{(j,j);\bm k}|^2)$. Then, define $I_m = [\log_2 q_{m-1}, \log_2 q_{m})$ for $i = 1, ..., M-1$ and $I_M = [\log_2 q_{M-1}, \log_2 q_{M}]$. In the linear relationship, we will use $c_m$, the $\log_2$ of the midpoints of the wavelet energy quantiles: 
\begin{equation}
    c_m = \log_2 \left[(q_{m - 1} + q_m)/2 \right], \quad m = 1, \ldots, M.
\end{equation}

Now, we calculate $\bar j_m$, the average level of detail coefficients with $\log_2$ wavelet energies in the interval $I_m$. In the following equations, elements $(l, \bm k)$ of the set $A_m$ are such that $[(l, l), \bm k]$ is a valid index of a 2D detail wavelet coefficient. For the 2D NDWT of a square image with side length $N = 2^J$, then $l \in \{0, 1, ..., J-1 \}$ and $\bm k = (k_1, k_2) \in \{1, 2, ..., N\}^2$.

\begin{equation}
    \bar j_m = \frac{1}{|A_m|} \sum_{(j, \bm k) \in A_m} j, \quad A_m = \{(l, \bm k): \log_2|d_{(l, l); \bm k)}|^2 \in I_m \}.
\end{equation}

This average can equivalently be rewritten as a weighted average of levels, where the weights are relative frequencies of levels for detail coefficients with $\log_2$ wavelet energies in $I_m$:

\begin{equation}
    \bar j_m = \sum_{j=J-L}^{J-1} j \times w_j, \text{ where }
    w_j = \frac{| A_m \cap \{ (l, \bm k) : l = j \} |} {|A_m|}.
\end{equation}

In this way, $\bar j_m$ can be interpreted as an estimate of the conditional expectation $\mathbb E (j | \log_2 |d_{(l,l);\bm k}|^2 \in I_m)$, where $w_j$ is an estimate of $P (l = j | \log_2 |d_{(l,l);\bm k}|^2 \in I_m)$. Although it may seem unusual to consider the level indices as random, this observation fits with the analogy of the inverted primal-dual relationship: just as the primal spectra considers the $\log_2$ wavelet energies as random for a fixed level $j$, the dual spectra is effectively treating the levels as random, for fixed $\log_2$ wavelet energies.

The dual spectra equation with these quantities is
\begin{equation}\label{eq:dualequation}
\bar j_m = -[1 / (2H + 2)] c_m + C'_{\psi, H}.    
\end{equation}
Let $\beta_d = -1 / (2H + 2)$ be the slope of the dual spectra, and let $\hat \beta_d$ be an estimate of this slope found by plotting points $(\bar j_m, c_m)$. The dual spectra estimator for $H$ is 
\begin{equation}\label{eq:dualHest}
\hat H_d = - \frac 1 2 \left(\frac {1} {\hat \beta_d} + 2 \right).
\end{equation}

\begin{figure}[h!]
\centering
\begin{subfigure}{0.4\textwidth}
    \includegraphics[width=\textwidth]{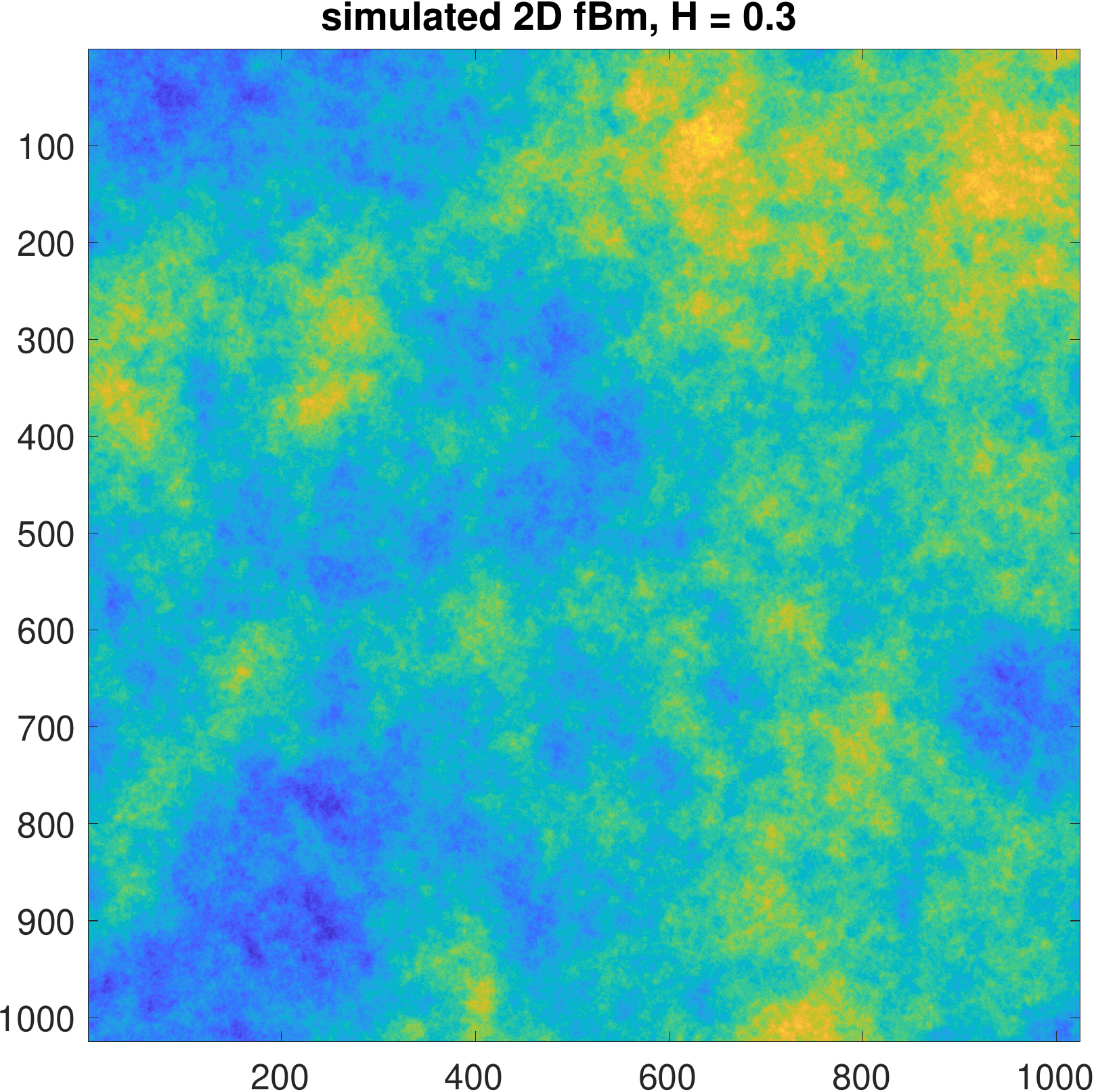}
    \caption{}
    \label{fig:dual-schem-1}
\end{subfigure}
\hspace{2cm}
\begin{subfigure}{0.4\textwidth}
    \includegraphics[width=\textwidth]{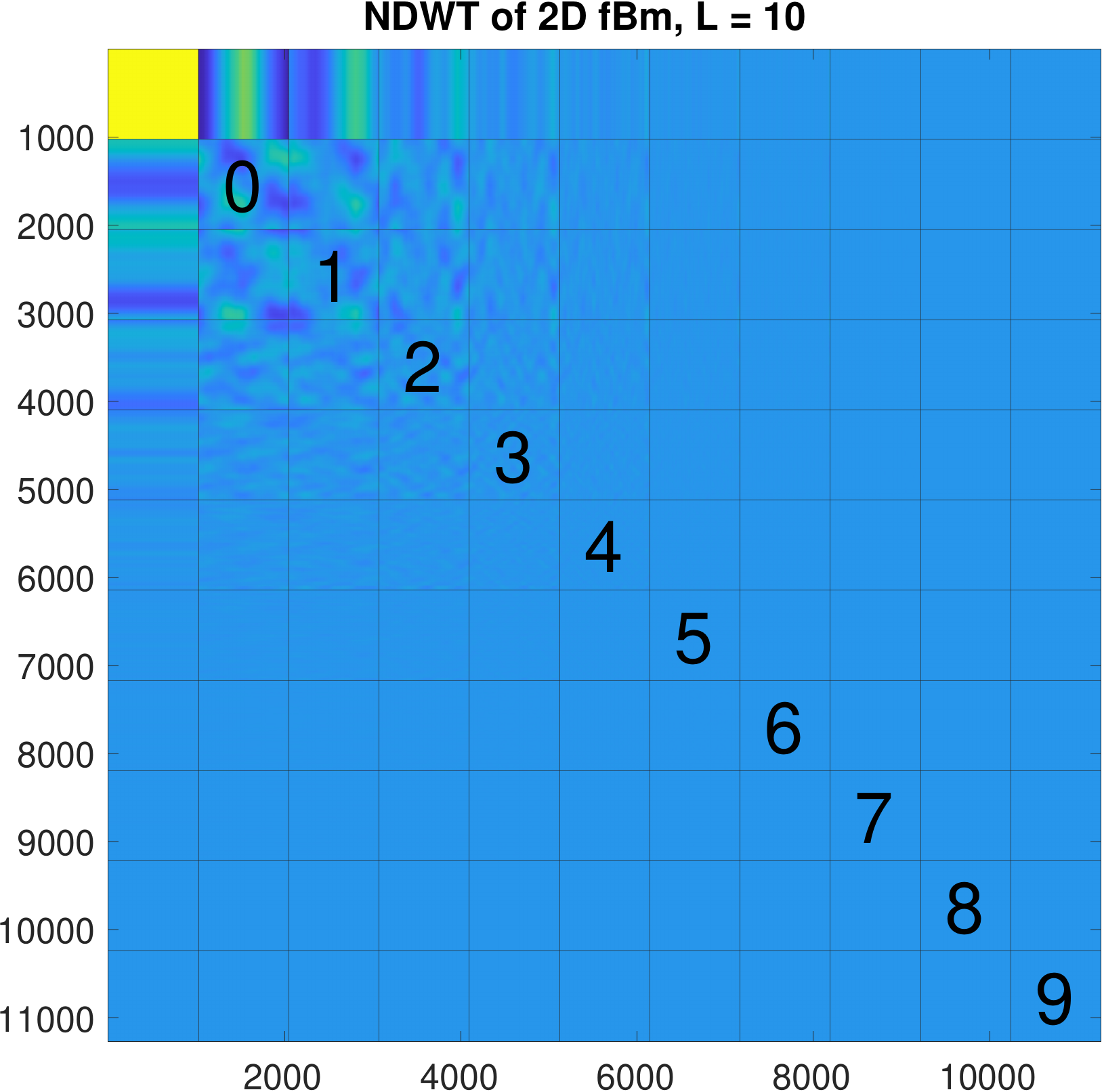}
    \caption{}
    \label{fig:dual-schem-2}
\end{subfigure}
% \hfill
\hspace{2cm}
\begin{subfigure}{0.4\textwidth}
    \includegraphics[width=\textwidth]{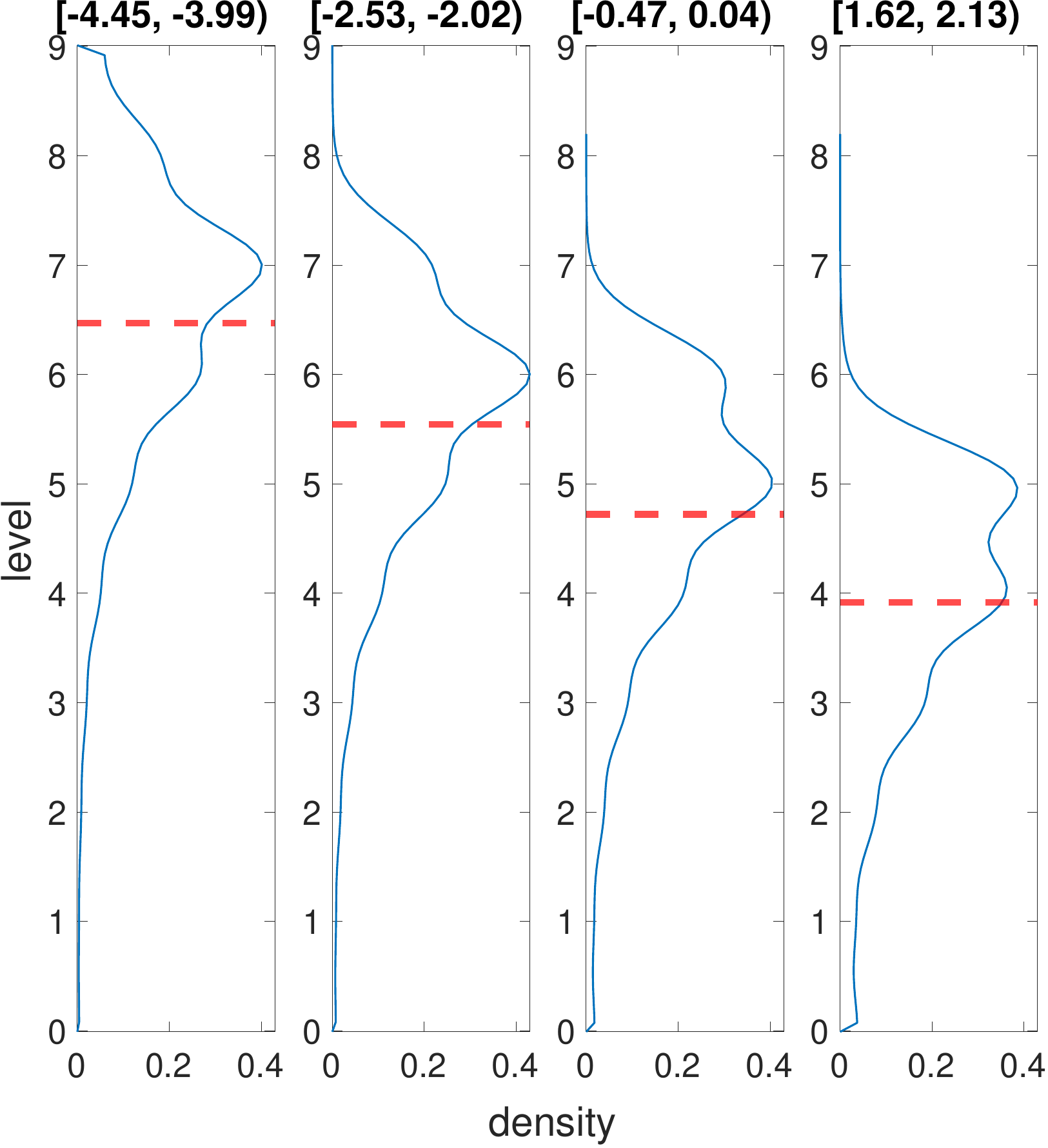}
    \caption{}
    \label{fig:dual-schem-3}
\end{subfigure}
\hspace{2cm}
\begin{subfigure}{0.4\textwidth}
    \includegraphics[width=\textwidth]{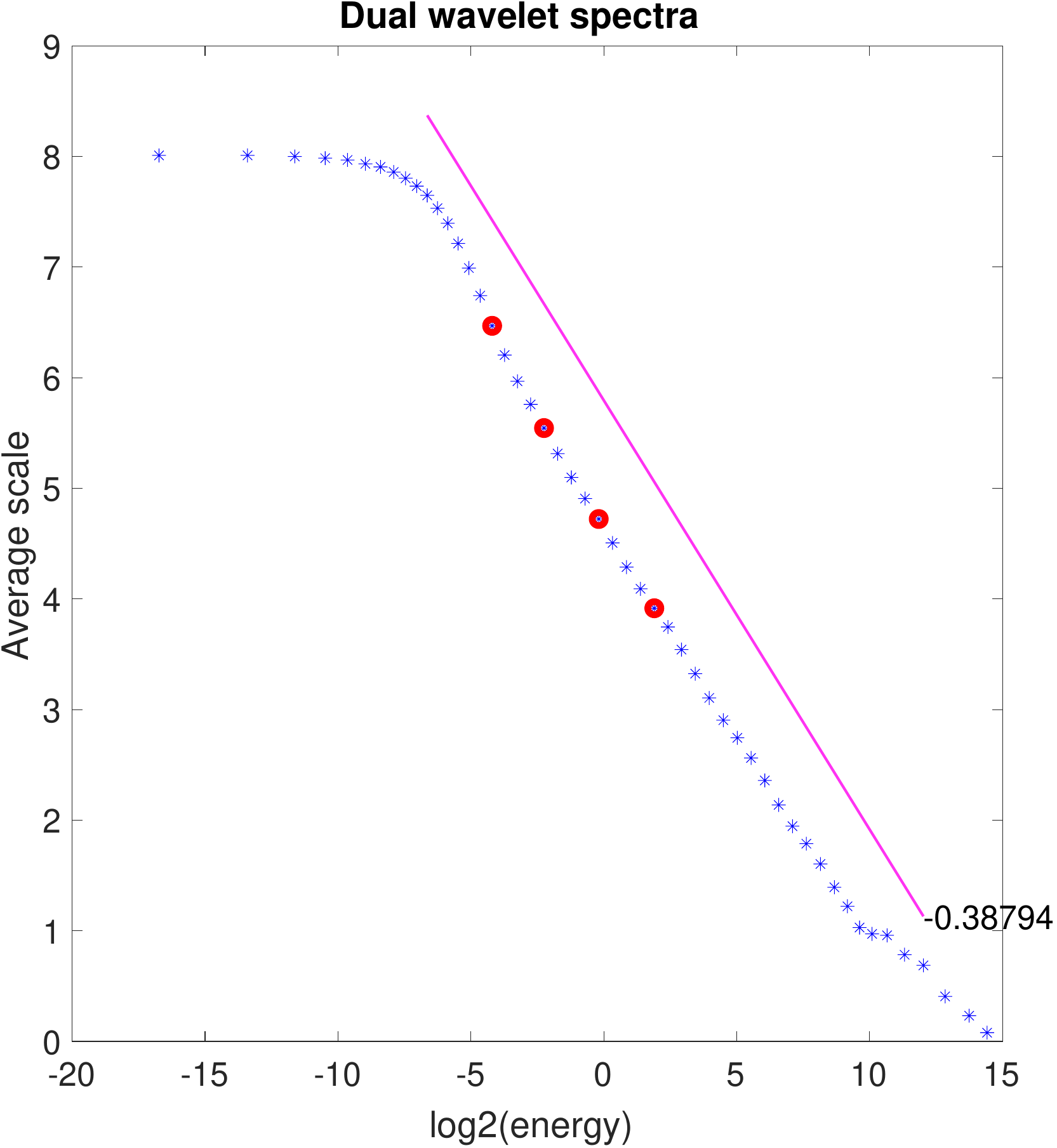}
    \caption{}
    \label{fig:dual-schem-4}
\end{subfigure}
        
\caption{Schematic of the scale-mixing 2D NDWT and the distribution of levels for four example intervals of log-wavelet energies. (a) Simulated $1024 \times 1024$ 2D fBm signal with $H = 0.3.$ (b) Scale-mixing NDWT with $L = 10$ details levels. Numbers indicate levels of diagonal detail coefficients. Each region is $1024 \times 1024.$ (c) Distribution of levels for four selected wavelet energy intervals. Red dashed lines indicate the average level. (d) dual spectra, with a line fit using the 0.2 to 0.95 quantiles. Red markers show the selected intervals from panel (c). The slope is $\hat \beta_d = -0.38794,$ leading to the estimate $\hat H_d = 0.2889.$}
\label{fig:NDWT-dual-schematic}
\end{figure}

Figure \ref{fig:NDWT-dual-schematic} contains an example of the steps involved in computing the dual spectra. Figure \ref{fig:dual-schem-1} is a simulated 2D fractional Brownian motion, and Figure \ref{fig:dual-schem-2} shows the 2D NDWT with $L = 10$ detail levels. In Figure \ref{fig:dual-schem-3}, we show the estimated density of levels for detail coefficients in four intervals of $\log_2$ wavelet energies. In practice, we do not estimate these densities for the dual spectra, but only compute the sample average levels, shown in dashed red lines. However, this figure emphasizes the dual perspective of considering the levels $j$ to be random, given that the corresponding detail coefficients have $\log_2$ wavelet  energies in a specific interval. Finally, Figure \ref{fig:dual-schem-4} shows the plot of dual spectra points, with a line fit using the 0.2 to 0.95 quantiles of the $\log_2$ wavelet energy values. The slope of the line is $\hat \beta_d = -0.38794,$ and thus the estimated Hurst exponent is $\hat H_d = 0.2889.$

For a theoretical self-similar Gaussian process, the estimates from the primal and dual spectra target the same parameter, $H$. However, for general scaling processes, primal and dual spectra may carry complementary information, and thus estimates from both approaches may summarize different characteristics. As an example of such a phenomenon, we simulated signals with simultaneous characteristics of high and low Hurst exponent values, and compared the estimation of $H$ with the primal and dual spectra. Specifically, we simulated separate 2D fBm with $H = 0.4$ and $H = 0.6$, performed the 2D NDWT, and replaced the finest detail levels of the $H = 0.6$ signal with those from the $H = 0.4$ signal. We then inverted the 2D NDWT, yielding an image with mixed $H$ characteristics. We repeated this process, and estimated $H$ for all signals with the primal and dual spectra. A paired $t$-test showed that the estimates from the primal and dual spectra were significantly different ($p$-value $\approx 0$). That is, in the presence of non-ideal self-similarity, estimates from the primal and dual spectra did not coincide, demonstrating their potential to provide complementary information. In our application, we will use estimates of $H$ from both approaches as features for classifying real data, to see whether combining these features leads to improved classification performance.

\subsection{Implementation remarks}\label{sec:implement}

The number of intervals, $M,$ or quantiles, $n_q = M - 1$, must be chosen. This number should be balanced to allow for a large number of observations within each interval and a sufficient number of points to construct the regression line. We will specify $n_q$ relative to the number of detail levels in the decomposition. For example, with $L = 10$ detail levels, we may consider multiples 2, 5, and 10, or $n_q \in \{2L, 5L, 10L\} = \{20, 50, 100 \}.$ In the simulations and application, this multiple is referred to as \texttt{xq} or \texttt{xquantile}.

After plotting the dual spectra, one can also choose to use only a subset of the points for estimating the slope. This is analogous to the primal spectra, when the spectra is often only linear for a subset of levels, especially for real-world datasets. For example, we may specify that we will use dual spectra points with $\log_2$ wavelet energies in the 0.1 to 0.85 quantiles. These are referred to as \texttt{q1, q2} (or as percentiles, \texttt{p1, p2}). The dual spectra typically deviate from linearity most severely at extreme quantiles, so a central range is likely to avoid such regions. However, a small number of quantiles reduces the number of points available to estimate the slope, possibly leading to less stable estimates.

Although the derivation of the dual spectra applies to both the 2D DWT and NDWT, we recommend using the NDWT. At a high level, this is because the NDWT leads to more data points than the classic DWT, where the number of coefficients is halved at each level. For more intuition about why these additional data points are desirable, consider again that the weight $w_l$ can be seen as an estimate of a probability. Using the DWT, some of these weights will be exactly 0 if there are fewer coefficients in a level than there are intervals in the dual spectra. For example, in the coarsest detail levels of the DWT, there will only be 1, 4, 16, or 64 diagonal detail coefficients. If we use 50 quantiles, or 51 intervals, then $w_l$ will be exactly 0 for $l = 0, 1, 2$ in most intervals. By contrast, for a $1024 \times 1024$ image, the coarsest detail levels in the NDWT will all contain $1024^2$ diagonal detail coefficients, allowing a more stable estimate of these probabilities.

Finally, in our implementation, we use the scale-mixing 2D NDWT because it can be calculated with a simple matrix multiplication. However, the standard 2D NDWT can also be used, because the diagonal region detail coefficients are the same in both methods.

\subsection{Simulation study} \label{sec:sim-study}

A simulation study was carried out to learn how the estimator performs with different settings, and to compare its performance to the primal spectra-based estimators. Simulated 2D fBm of size $1024 \times 1024$ were generated, for known values of $H \in \{0.1, 0.2, \ldots, 0.9 \}$. The DWT and NDWT were calculated with 10 detail levels using a chosen wavelet. The dual spectra method was used to calculate $\hat H_d$ with a chosen combination of number of quantiles and range of quantiles. The primal spectra method was used to calculate $\hat H_{p,DWT}$ and $\hat H_{p,NDWT}$, where levels $j_1=2$ to $j_2=8$ were used for estimating the primal spectra slopes.

\begin{figure}[t]
    \centering
    \includegraphics[width=1\linewidth]{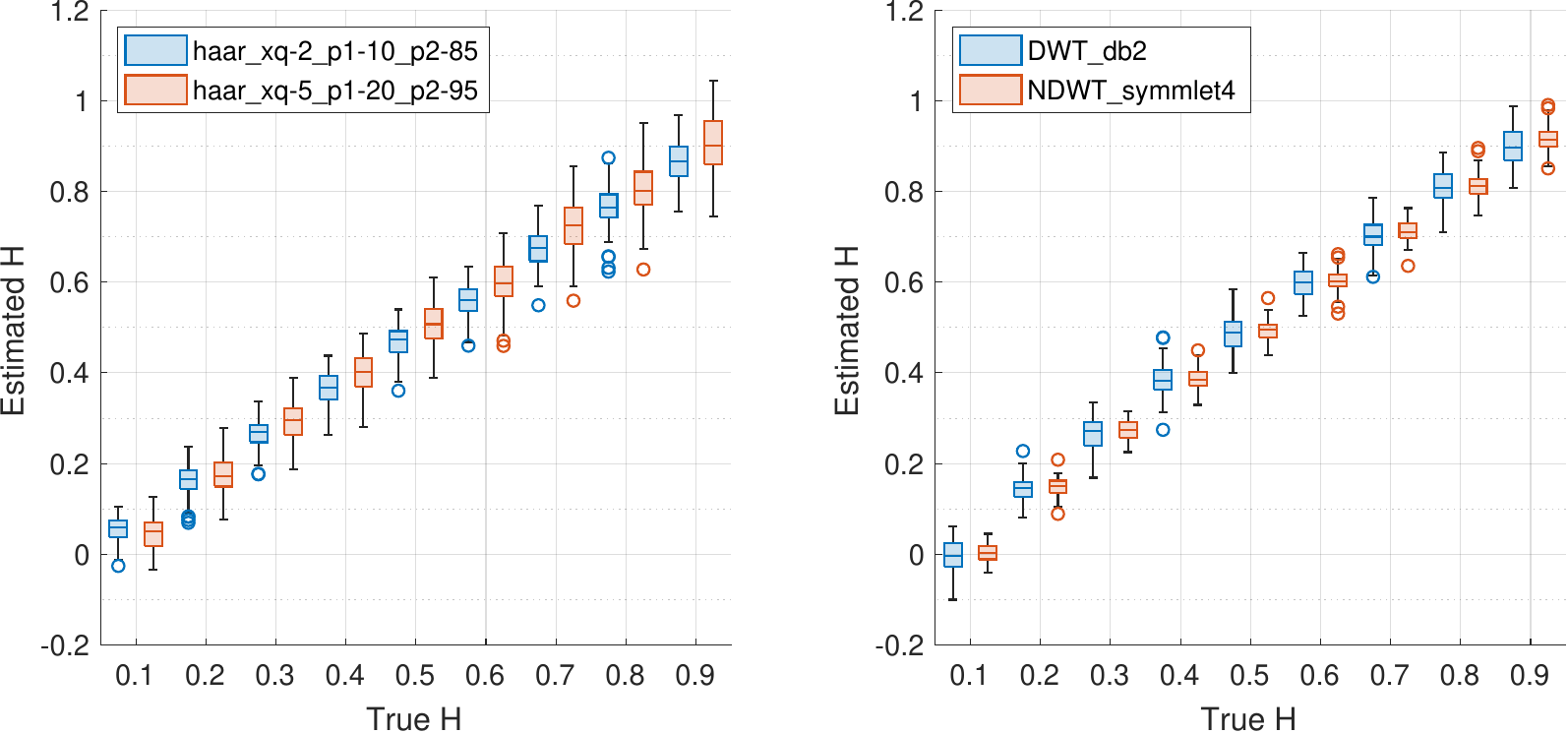}
    \caption{\small Comparison of estimates of $H$ by dual (left) and primal (right) wavelet spectra. Legend labels refer to settings described in Section \ref{sec:implement}.}
    \label{fig:sim-image}
\end{figure}

For each $H$ value, $N=100$ simulated signals were generated. The following wavelets were used: Haar, Daubechies with 2 and 3 vanishing moments, Pollen wavelet with one argument ($\phi = \pi/4$), Coiflet wavelet with 6 filter taps, Symmlet with 4 vanishing moments, and symmetric complex orthonormal wavelet \citep{sherlock2002matlab} with 6 filter taps (respectively, haar, db2, db3, pollen, coif1, symmlet4, and conf6). For the dual spectra method, the numbers of quantiles used were $\{20, 50, 100\}$. The quantile ranges $[p_1,p_2]$ used were chosen from $\{[10, 85], [20, 95], [15, 65], \allowbreak [25, 75], [35, 85], [45, 95]\}$. The squared error was computed for each simulated signal and estimate, and the average of mean squared errors (AMSE) across all $H$ values was computed to compare the overall performance of the settings. For clarity, define $AMSE(\hat H) = \frac{1}{9000}\sum_{H \in \{0.1, ..., 0.9 \}} \sum_{i = 1}^{100} (\hat H_i - H)^2.$ The four best (lowest AMSE) parameter settings from each method are summarized in Table \ref{tab:sim-results}. From the table it can be seen that these four settings have similar performance in estimating $H$, and their AMSEs are on a similar order of magnitude to the primal DWT and NDWT estimators. The performance of a selection of those settings at different $H$ values is visualized in Figure \ref{fig:sim-image}. It can be seen that estimates from primal spectra tend to have lower variability, but estimates from dual spectra are less biased for some $H$, especially $H = 0.1$ and $0.2$.

\begin{table}[t]
\centering
\caption{Summary of $H$ estimation in simulations with best settings according to AMSE}
\label{tab:sim-results}
\begin{tabular}{lllr} 
\toprule
\textbf{Method} & \textbf{Wavelet} & \textbf{Settings}          & \multicolumn{1}{l}{\textbf{AMSE}}  \\
\midrule
dual            & haar             & xquantile=2,\ $p_1$=10, $p_2$=85  & 0.002656 \\
dual            & haar             & xquantile=5,\ $p_1$=20, $p_2$=95  & 0.003088 \\
dual            & haar             & xquantile=10,\ $p_1$=20, $p_2$=95 & 0.003113 \\
dual            & haar             & xquantile=5,\ $p_1$=10, $p_2$=85  & 0.003198 \\
\midrule
primal, NDWT    & symmlet4         & $j_1=2$, $j_2=8$                 & 0.002093 \\
primal, NDWT    & conf6            & $j_1=2$, $j_2=8$                & 0.002102 \\
primal, NDWT    & db3              & $j_1=2$, $j_2=8$                 & 0.002102 \\
primal, NDWT    & coif1            & $j_1=2$, $j_2=8$                 & 0.002114 \\
\midrule
primal, DWT     & db2              & $j_1=2$, $j_2=8$                 & 0.002984 \\
primal, DWT     & coif1            & $j_1=2$, $j_2=8$                 & 0.003064 \\
primal, DWT     & db2              & $j_1=2$, $j_2=8$                 & 0.003131 \\
primal, DWT     & db2              & $j_1=2$, $j_2=8$                 & 0.003262 \\
\bottomrule
\end{tabular}
\caption*{\small{Entries in the Settings column are described in Sections \ref{sec:self-sim}} and \ref{sec:implement}.}
\end{table}

The best dual estimator had a larger AMSE than the best primal NDWT-based estimator, and a smaller AMSE than the best primal DWT-based estimator, as seen in Table \ref{tab:sim-results}. This suggests that the primal NDWT-based estimator may perform better than the dual spectra-based estimator across all $H$ values on average. For applications, though, it is also helpful to understand how the estimators perform at different values of $H$. Figure \ref{fig:sim-image} provides a visual summary of the center and spread of the estimates at the tested values of $H$. Between the two dual estimators, the setting with 20 quantiles using quantiles in $[0.1, 0.85]$ produced estimates that were less variable than those with 50 quantiles using quantiles $[0.2, 0.95]$. On the other hand, the latter settings are often less biased, especially at higher $H$ values. In both cases, the variability of the estimates increases slightly as $H$ increases.

Compared to the primal $H$ estimates, the dual $H$ estimates have larger variability than those from the primal NDWT estimator. The dual estimators are considerably less biased than the primal estimators for $H=0.1$, and slightly less biased for $H=0.2$. In summary, two settings for the dual estimator resulted in similar performance in terms of AMSE, although they exhibit different performance in terms of their bias and variability at different $H$ values. The best primal NDWT estimator performed better than the best dual estimators in terms of variability at individual $H$ values and AMSE across all $H$. However, the dual estimators are less biased at low $H$.

Across the 126 combinations that were tested for the dual spectra estimator, 8 of the top 10 best performing settings by AMSE used the Haar wavelet. All of the top 6 best performing settings by AMSE used the Haar wavelet with either \texttt{p1=10, p2=85} or \texttt{p1=20, p2=95}. The other two combinations in the top 10 used the Coiflet wavelet with 6 filter taps. The largest AMSE of the top 10 settings was 0.004243. Thus, we recommend using the Haar wavelet and either \texttt{(p1, p2)=(10, 85)} or \texttt{(20, 95)}.

As noted, the Haar wavelet yielded the best performance among the tested wavelets. To evaluate the behavior of wavelets with larger support, such as those from the Daubechies and Symmlet families, we conducted numerical experiments and observed that when the wavelet filter extends beyond the signal boundaries, boundary effects introduce large coefficients that distort the distribution of levels $j$. This issue could be mitigated by considering only those coefficients generated when the wavelet filter is fully contained within the signal domain, or by applying mirror or constant extension to the data. Under such scenarios, a wavelet other than Haar might become optimal, potentially improving the accuracy of $H$ estimation. Another possibility is to consider a robust estimation of the mean of the $j$’s within a particular interval of $\log_2$ wavelet energies.

\section{Application to mammogram data/image classification} \label{sec:application}

As an example of an application, we apply this method to extract features from mammograms and classify whether the patient has breast cancer. We analyze the images with minimal preprocessing and do not specifically focus on regions of interest identified by experts. This highlights the ability of self-similarity quantification to identify complex patterns due to the presence or lack of pathologies. The goal for this application is to illustrate how the inclusion of information from the dual spectra enhances the classification performance. In this section, we describe the dataset used for this study, how features were extracted from the mammograms, and how these features were used to classify the patients. 

\subsection{Description of dataset and preprocessing}

The mammogram images are from the University of South Florida Digital Database for Screening Mammography (USF DDSM) \citep{heath1998current, heath2001digital}, as compiled by \cite{lekamlage2020miniddsm}. This dataset has been widely used as a benchmark in the scientific community because of its free accessibility and large, diverse collection of cases. Notably, a number of papers dealing with wavelet-based tools have considered this collection of images in their analyses. See for instance \cite{Nicolis2011, ramirez2013_2d_wavelet_multiscale,roberts2017wavelet,oyelade2022novel, kang2022NDWT2dspectrum}. However, our image pre-processing steps are deliberately simpler than those in related studies, to illustrate the effectiveness of our method even with non-ideal data. Due to this difference, our results are not directly comparable with these papers.

The DDSM comprises 2620 mammogram studies labeled as Normal, Cancer, or Benign collected with four different scanners at three locations. The labels were determined through mammogram readings, follow-up exams, and/or tissue biopsies as described in \cite{heath2001digital}. To control for the effect of the scanner and location, we chose one combination which yielded the largest sample size: subset A (Massachusetts General Hospital, MGH) with the HOWTEK scanner. Each study contains mediolateral oblique (MLO) and craniocaudal (CC) projections of the left and right breasts (four total). In our analysis, we only used the CC projections. The left or right image with the suspicious region was chosen for Cancer and Benign cases, and one left or right image was chosen randomly for Normal cases. The resulting sample contained 216 observations, with 63 Normal, 66 Cancer, and 87 Benign cases. We focused on classifying Normal and Cancer patients, and excluded Benign patients, for a final sample size of 129 observations.

\begin{figure}[t!]
\centering
\begin{subfigure}[c]{0.4\textwidth}
    \includegraphics[width=\textwidth]{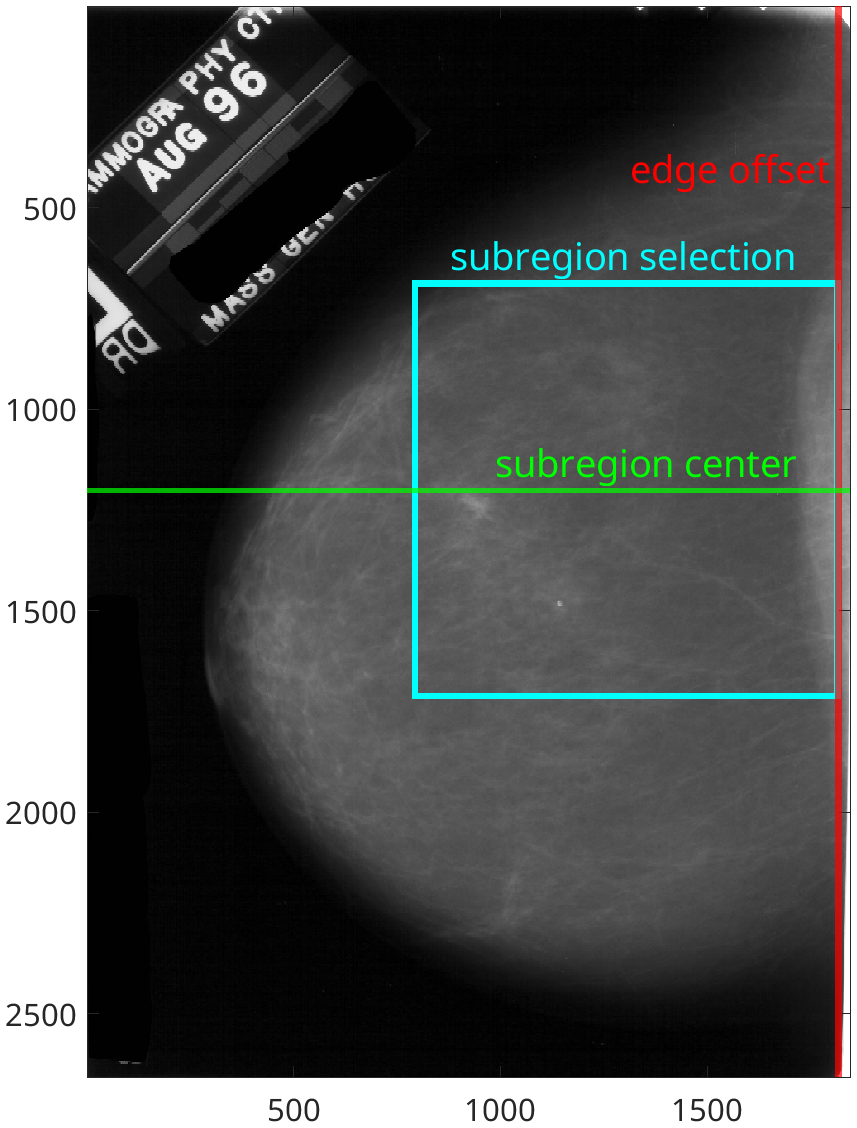}
    % \caption{}
    % \label{fig:preprocess-full}
\end{subfigure}
\begin{subfigure}[c]{0.4\textwidth}
    \includegraphics[width=\textwidth]{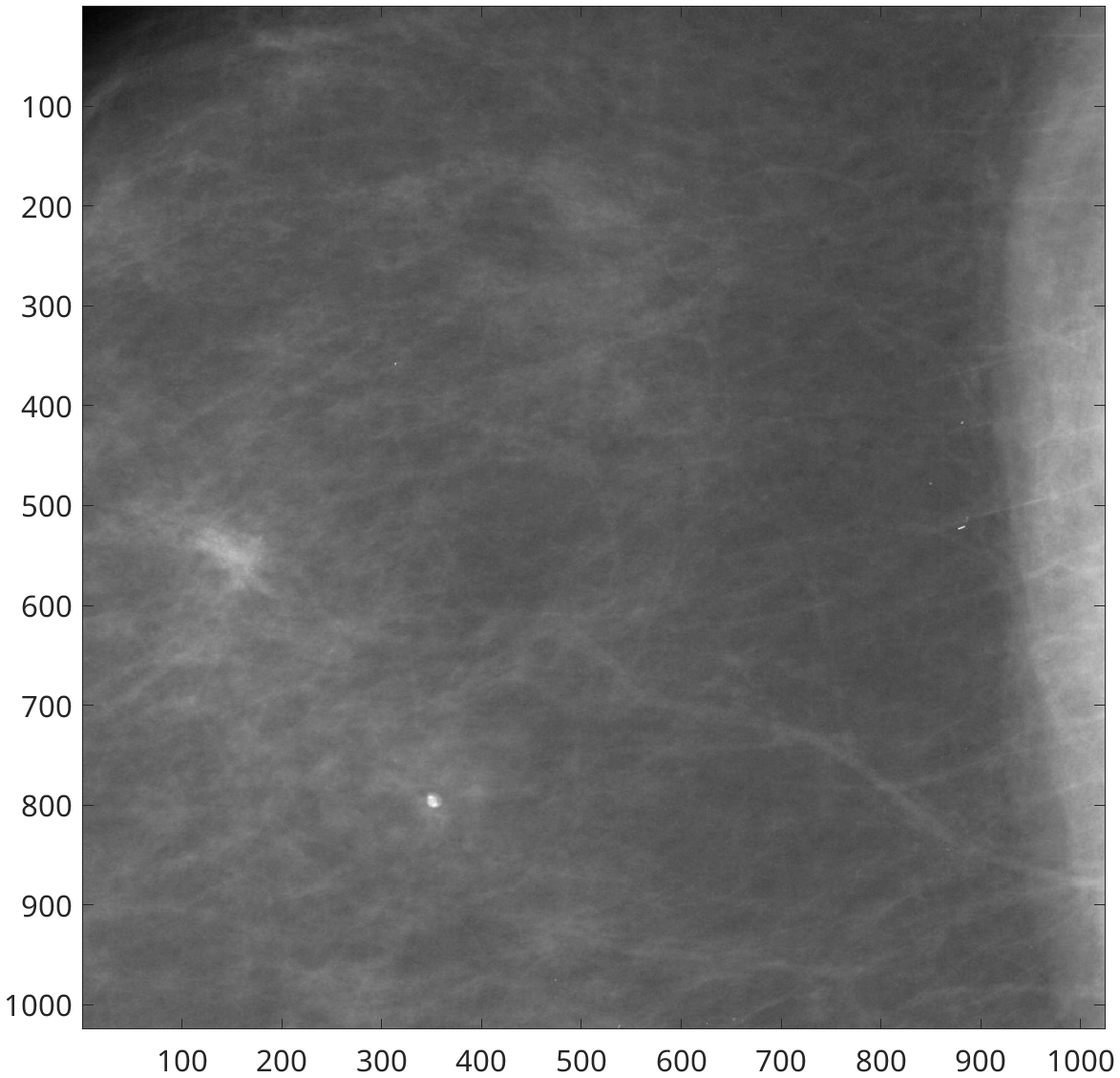}
    % \caption{}
    % \label{fig:preprocess-sub}
\end{subfigure}

\caption{Left: example of a full mammogram image from the DDSM and a random region selected to minimize inclusions of the background. Right: the selected region used for classification.}
\label{fig:mammo-schematic}
\end{figure}

It is important to point out that the final sample of images analyzed in this paper is not the same as those in the previously referenced papers. In particular, our preprocessing steps are relatively simple compared to other works, which may use multiple regions per mammogram, rely on the expert-identified regions-of-interest (ROI) annotations that are included with the dataset, or produce masks of the images to exclude non-tissue pixels, among other strategies. Our goal during preprocessing was to extract the largest region of the image with minimal background present and with dimensions that are a power of 2. This led to selecting regions of size $1024 \times 1024$. Depending on the orientation of the breast, the opposite region edge was fixed to the opposite image edge with an offset of 30 pixels (e.g. for a mammogram with the breast oriented to the right, the leftmost edge of the region was fixed to column 30 of the original image, where the column indices start at 0). The vertical position of the region was selected by randomly choosing a location for the center of the vertical edge. If the random location was 512 or more pixels away from the top and bottom of the image, then the selected value was used as the vertical center. Otherwise, a region extending 1024 pixels from the top or bottom of the image was used, depending on which side the random location was closest to. An example is shown in Figure \ref{fig:mammo-schematic}. Scripts to perform these preprocessing steps are available in a \href{https://gitlab.com/rayhinton/dual_wavelet_spectra/-/tree/main/datapreprocessing?ref_type=heads}{GitLab repository} \citep{Hinton2025gitdual}. 

\subsection{Feature extraction, classification, and evaluation}\label{sec:featX-class-eval}

Feature extraction was performed by estimating $H$ for each of the resulting images using the best settings by AMSE as found in the simulation study. For the dual spectra method, the two best setting combinations were used, to compare their performance. The entropy, denoted $E$, of the finest (first) level of detail coefficients was also calculated, as an example of an additional feature which summarizes different characteristics of the images and may complement features estimating $H$. This is based on the multiresolution wavelet entropy \cite{labat2005mwe}, which measures the amount of disorder in the detail wavelet energies at a particular level. The following combinations of features were used: individual $H$ features; combined primal and dual $H$ features; and primal $H$ and $E$ features with and without dual $H$ features. Standard logistic regression was used to predict the probability of patients having cancer or not (1 or 0), and a probability threshold of 0.5 was used to classify observations. Performance on unseen data was evaluated using 10 repetitions of stratified 10-fold cross-validation (CV). The average and standard deviation of sensitivity, specificity, and accuracy were calculated and are shown in Table \ref{tab:classif-metrics}. 

We use two approaches to understand the effect of including dual $H$ features in the model. First, we used the corrected repeated $k$-fold CV test from \cite{bouckaert2004hypotests} to check for statistically significant improvements in classification accuracy. This test controls the Type I error of the standard paired sample $t$-test by accounting for the dependence between the training and test sets of repeated $k$-fold CV. Table \ref{tab:classif-t-test} contains one-sided test $p$-values comparing feature sets with and without dual spectra features, where the null hypothesis is that the feature sets had the same accuracy, and the alternative hypothesis is that the feature sets with dual spectra features had higher accuracy.

Second, we calculated the difference in deviances for pairs of models to check for a significant effect of dual $H$ features on the log-odds of a Cancer diagnosis. Specifically, we calculated $D = D_1 - D_2$ by fitting logistic regression models $M_1$ and $M_2$ with the entire data set, where $M_1$ is a set of features with no dual $H$ feature, and $M_2$ has a dual $H$ feature added. Since only one feature is added, this test statistic has an approximate $\chi_1^2$ distribution \cite{mccullagh1989glms}. If we let $\beta_{\text{dual}}$ be the coefficient for some dual $H$ feature, then we test $H_0: \beta_{\text{dual}} = 0$ vs. $H_A: \beta_{\text{dual}} \neq 0$. The $p$-value for this test is $1 - F(D)$, where $F$ is the cumulative distribution function for a $\chi_1^2$ random variable. See Table \ref{tab:deviance-chi-sq-test} for the results of these tests.

\begin{table}[t]
\centering
\caption{Average classification metrics ($\pm$ standard deviation) from 10 repetitions of 10-fold CV}
\label{tab:classif-metrics}
\begin{tabular}{p{3cm}p{4cm}ccc}
\toprule
\textbf{Description} & \textbf{Feature set} & \textbf{Sensitivity} & \textbf{Specificity} & \textbf{Accuracy} \\
\midrule
\multirow{4}{=}{single $H$ features}                                          & $\hat H_{p,DWT}$ & $0.6500 \pm .1929$     & $0.5493 \pm .1959$     & $0.5997 \pm .1216$   \\
& $\hat H_{p,NDWT}$ & $0.6700 \pm .1917$     & $0.5079 \pm .1811$     & $0.5905 \pm .1155$   \\
& $\hat H_d$ & $0.7071 \pm .1818$     & $0.5236 \pm .1823$     & $0.6184 \pm .1162$    \\
& $\hat H_{d2}$ & $\textbf{0.7260} \pm .1723$     & $\textbf{0.5648} \pm .1908$     & $\textbf{0.6486} \pm .1154$ \\
\midrule
\multirow{4}{=}{add dual $H$ features to primal $H$ features} & $\hat H_{p,DWT}, \hat H_d$ & $0.6969 \pm .1808$     & $0.5814 \pm .1902$     & $0.6403 \pm .1272$ \\
& $\hat H_{p,DWT}, \hat H_{d2}$ & $\textbf{0.7364} \pm .1630$    & $\textbf{0.6250} \pm .1568$     & $\textbf{0.6812} \pm .1067$ \\
& $\hat H_{p,NDWT}, \hat H_d$ & $0.7171 \pm .1937$     & $0.5771 \pm .1774$     & $0.6482 \pm .1269$ \\
& $\hat H_{p,NDWT}, \hat H_{d2}$ & $0.7269 \pm .1672$     & $0.6081 \pm .1672$     & $0.6686 \pm .1121$ \\
\midrule
\multirow{6}{=}{add dual $H$ feature to primal NDWT $H$ and entropy features} & $\hat H_{p,DWT}, E_{p,DWT}$           & $0.6486 \pm .2007$     & $0.4998 \pm .1794$     & $0.5756 \pm .1225$ \\
& $\hat H_{p,DWT}, E_{p,DWT}, \hat H_d$      & $0.6695 \pm .1785$     & $0.5691 \pm .1983$     & $0.6202 \pm .1283$ \\
& $\hat H_{p,DWT}, E_{p,DWT}, \hat H_{d2}$   & $0.7029 \pm .1698$     & $0.6138 \pm .1739$     & $0.6585 \pm .1101$ \\
& $\hat H_{p,NDWT}, E_{p,NDWT}$         & $0.6291 \pm .1884$     & $0.5129 \pm .1891$     & $0.5718 \pm .1196$ \\
& $\hat H_{p,NDWT}, E_{p,NDWT}, \hat H_d$    & $0.7171 \pm .1748$     & $0.6031 \pm .1833$     & $0.6605 \pm .1223$ \\
& $\hat H_{p,NDWT}, E_{p,NDWT}, \hat H_{d2}$ & $\textbf{0.7817}{^\ast} \pm .1515$     & $\textbf{0.6305}{^\ast} \pm .1798$     & $\textbf{0.7072}{^\ast} \pm .1200$\\
\bottomrule
\end{tabular}
\caption*{\footnotesize{Features calculated with the respective settings that had lowest AMSE in simulations. $H$ = Hurst exponent; $E$ = wavelet entropy from finest detail level; $d$ = dual setting with lowest AMSE; $d2$ = dual setting with 2nd lowest AMSE; $p, NDWT$ = primal NDWT settings with lowest AMSE; $p, DWT$ = primal DWT settings with lowest AMSE. \textbf{Bold} indicates best (highest) metric value within a group of feature sets. Asterisk${^\ast}$ indicates best metric value overall.}}
\end{table}

\begin{table}[t]
\centering
\caption{Hypothesis test $p$-values for differences in average accuracies}
\label{tab:classif-t-test}
\begin{tabular}{llll}
                    & & \multicolumn{2}{c}{$B$}                                           \\
\toprule
\multirow{13}{*}{$A$} & $A$ vs. $B$     & $\hat H_{p,DWT}$              & $\hat H_{p,NDWT}$           \\
\cmidrule{2-4}
                    & $\hat H_d$       & 0.3147           & 0.1950         \\
                    & $\hat H_{d2}$    & 0.08398          & 0.04773$^\ast$       \\
\cmidrule{2-4}
                    & $(A, B)$ vs. $B$ & $\hat H_{p,DWT}$              & $\hat H_{p,NDWT}$           \\
\cmidrule{2-4}
                    & $\hat H_d$       & 0.2501           & 0.1515                \\
                    & $\hat H_{d2}$    & 0.07344          & 0.08892               \\
\cmidrule{2-4}
                    & $(A, B)$ vs. $B$ & $\hat H_{p,DWT}, E_{p,DWT}$ & $\hat H_{p,NDWT}, E_{p,NDWT}$ \\
\cmidrule{2-4}
                    & $\hat H_d$       & 0.2098          & 0.05727                 \\
                    & $\hat H_{d2}$    & 0.07167         & 0.01245$^\ast$            \\
\bottomrule
\end{tabular}
\caption*{\footnotesize{Hypothesis test $p$-values comparing average accuracies using the corrected repeated $k$-fold CV test from \cite{bouckaert2004hypotests}. Let $F_X$ and $F_Y$ be the accuracies of feature sets $X$ vs. $Y$. The table contains $p$-values for one-sided tests: $H_0: F_X = F_Y$ vs. $H_1: F_X > F_Y$. Asterisk$^\ast$ indicates significance at the 5\% level.}}
\end{table}

\begin{table}[t]
\centering
\caption{Hypothesis test $p$-values for significance of dual $H$ feature effects}
\label{tab:deviance-chi-sq-test}
\begin{tabular}{llll}
                    & & \multicolumn{2}{c}{$B$}                                           \\
\toprule
\multirow{8}{*}{$A$} & $B$ vs. $(A, B)$ & $\hat H_{p,DWT}$              & $\hat H_{p,NDWT}$           \\
\cmidrule{2-4}
                    & $\hat H_d$       & $2.439 \times 10^{-8}$ & $1.838 \times 10^{-9}$ \\
                    & $\hat H_{d2}$    & $1.277 \times 10^{-9}$ & $6.889 \times 10^{-10}$\\
\cmidrule{2-4}
                    & $B$ vs. $(A, B)$ & $\hat H_{p,DWT}, E_{p,DWT}$ & $\hat H_{p,NDWT}, E_{p,NDWT}$ \\
\cmidrule{2-4}
                    & $\hat H_d$       & $2.615 \times 10^{-8}$ & $6.333 \times 10^{-10}$ \\
                    & $\hat H_{d2}$    & $1.270 \times 10^{-9}$ & $1.999 \times 10^{-10}$ \\
\bottomrule
\end{tabular}
\caption*{\footnotesize{The hypothesis being tested is $H_0: \beta_{\text{dual}} = 0$ vs. $H_A: \beta_{\text{dual}} \neq 0$, where $\beta_{\text{dual}}$ is the coefficient for one of the dual $H$ features (notated as $A$ in the table). Reduced and full models were fit using the entire data set with feature sets $B$ and $(A, B)$, respectively. $p$-values are calculated assuming an approximate $\chi_1^2$ distribution for the difference in deviances.}}
\end{table}

\subsection{Discussion} \label{sec:discuss}

The estimated test metrics for feature sets with dual $H$ features are higher than the corresponding metrics for feature sets without such features. In particular, all of the best performing feature sets included the dual $H$ features calculated using the 2nd best simulation settings (denoted $\hat H_{d2}$). The best performance was achieved by combining the $\hat H_{d2}$ features with $H$ and $E$ calculated from best NDWT primal spectra settings. Work such as \cite{vimalajeewa2025multiscale} demonstrated the usefulness of combining primal wavelet spectra-based features with wavelet entropy in biomedical classification. Our results further show that dual spectra features lead to additional discriminative power when combined with both primal spectra features and entropy. Although the primal and dual spectra estimators estimate the same quantity for theoretical 2D fBm, they may extract complementary information in real data, and thus including features calculated from both spectra can improve classification performance.

The $p$-value for the test of a significantly higher accuracy with this feature set was 0.01245. However, care should be taken interpreting these $p$-values, since multiple hypotheses are being tested. A Bonferroni correction for a 0.05 Type I error rate across 12 tests would be $0.05/12 \approx .004167$, in which case we would fail to reject the null hypothesis. In practice, we observed that the $p$-value for this comparison was consistently below 0.05 when different random seeds were used. In Table \ref{tab:deviance-chi-sq-test}, we see that the corresponding $p$-value for this set of features is highly significant ($1.999 \times 10^{-10}$), and is much lower than the Bonferroni correction for eight simultaneous tests. Thus, there is evidence that the dual $H$ features have a significant effect on the log-odds of Cancer. However, the variability observed during cross-validation means that this does not translate to a significant increase in accuracy after adjustments for multiple testing.

We also note that the estimated accuracy of the best-performing features is low -- only 70.72\%. However, our primary objective in this example was not to maximize accuracy, but rather, to demonstrate the potential for combining dual and primal spectra features. Improved classification performance could be achieved by, for example, using more advanced pre-processing techniques, extracting more features, or using more complex machine learning algorithms. With logistic regression, improvements could be achieved with regularization, or by choosing the optimal probability threshold through analysis of the receiver operating characteristic curve. Although the improvement in accuracy was not found to be significant after controlling for multiple testing, dual spectra features were found to have a significant effect on the log-odds of Cancer, highlighting the capability of dual spectra $H$ estimates to complement those from the primal spectra in real-world data.

\section{Conclusion} \label{sec:concl}

In this paper, we have derived the empirical dual wavelet spectra as an inversion of the primal spectra, and shown how it can be used to quantify self-similarity in a signal by estimating the Hurst exponent. Although this new tool has been derived and demonstrated for the two-dimensional case, its extension to an arbitrary dimension is straightforward. Simulation studies provide guidance on the best parameters to use with this technique, and show that the dual spectra can be less biased than the primal spectra for some values of $H$. When used for classifying real world data, dual $H$ features had a statistically significant effect on the log-odds of predicting Cancer in mammograms, and models including these features achieved numerically higher average classification accuracy, sensitivity, and specificity, compared to models without dual $H$ features.

The concept of the dual spectra can be extended with other wavelet decompositions. For example, the continuous wavelet transform essentially decomposes a signal at a continuous range of levels, and thus may provide a finer grain estimate of the average level of detail coefficients with a given wavelet energy. Wavelet packets operate by also decomposing the detail coefficients at each level, which offers yet another perspective on the wavelet spectra relationship. When complex-valued wavelets are used, the distribution of levels vs. coefficient phase can be explored. It would also be interesting to use a robust estimate for the mean level $j$, such as the robust approaches in \cite{Hamilton2024waveletscaling} used for estimating the mean wavelet energy. Finally, it would be of interest to explore the application of the new tool in other contexts, not necessarily medical, but nonetheless involving data characterized by self-similarity, such as environmental or financial data, for instance, such as in \cite{ramirez-cobo2011wavelet} or \cite{ghosh2011characterizing}.

\paragraph{Acknowledgments.}
Raymond Hinton and Brani Vidakovic acknowledge the partial support of the H.O. Hartley Chair foundation and NSF Award 2515246 at Texas A\&M University. Pepa Ramírez-Cobo acknowledges research grants PID2022-137818OB-I00 (Ministerio de Ciencia e Innovación, Span), FEDER-UCA-2024-A2-35 and FQM-329 (Junta de Andalucía).

%	\newpage
\bibliography{refs}

% \appendix

% \section{Proofs}\label{app:1}
	
\end{document}